\documentclass{article}

\usepackage{arxiv}          

\usepackage[utf8]{inputenc} 
\usepackage[T1]{fontenc}    
\usepackage{graphicx}       
\usepackage{booktabs}       
\usepackage{amsmath}
\usepackage{microtype}
\usepackage{catchfile}      
\usepackage{authblk}        
\usepackage{orcidlink}      
\usepackage{xcolor}
\definecolor{linkblue}{rgb}{0.10,0.25,0.55}
\usepackage{hyperref}

\usepackage[style=authoryear,natbib=true,backend=biber,
            maxcitenames=2,mincitenames=1,
            uniquename=false,uniquelist=false]{biblatex}

\graphicspath{{../media/}}

\newenvironment{articlefigure}[1][htb]{\begin{figure}[#1]}{\end{figure}}
\newenvironment{articletable}[1][!ht]{\begin{table}[#1]\small}{\end{table}}
\newcommand{\backmattersection}[1]{\subsection*{#1}}

\title{TrunX: A Massively Parallel, Differentiable Implementation of the 3-PG Forest Growth Model in JAX}
\renewcommand{\shorttitle}{A JAX implementation of 3-PG}

\author[1]{\orcidlink{0000-0002-9344-8095}\,Glory Mary Givi\thanks{%
  Corresponding author: \href{mailto:glorymary.givi@hevs.ch}{\texttt{glorymary.givi@hevs.ch}}}}
\author[1]{\orcidlink{0009-0005-0364-6143}\,Cédric Travelletti}
\author[1]{\orcidlink{0000-0002-6027-7939}\,Grégory Mermoud}

\affil[1]{\parbox{0.9\textwidth}{\centering Institute of Informatics, School of
  Engineering, HES-SO Valais-Wallis University of Applied Sciences and Arts
  Western Switzerland, Rue de l'Industrie 23, 1950 Sion, Switzerland}}

\hypersetup{
  colorlinks=true, allcolors=linkblue,   
  pdftitle={TrunX: A Massively Parallel, Differentiable Implementation of the 3-PG Forest Growth Model in JAX},
  pdfauthor={Glory Mary Givi, Cédric Travelletti, Grégory Mermoud},
  pdfkeywords={3PG, automatic differentiation, differentiable modeling,
               forest growth model, JAX},
}

\begin{document}
\maketitle

\begin{abstract}
Process-based forest models are widely used to simulate forest growth and responses to
environmental change, but their calibration and application often require many
computationally expensive model evaluations. We present an implementation of the
Physiological Processes Predicting Growth (3-PG) model in JAX that uses just-in-time
compilation, vectorization, and GPU acceleration to reduce execution time. The
implementation also supports automatic differentiation, providing gradients of model
outputs and calibration objectives with respect to model parameters. This enables
efficient gradient-based optimization and gradient-informed Bayesian calibration,
extending 3-PG beyond conventional gradient-free approaches. The implementation produced
results numerically consistent with r3PG for the evaluated configuration. Overall, the
JAX implementation provides a faster and differentiable framework for calibrating and
applying the 3-PG model.

\end{abstract}


%
\newcommand{\preprintlicense}{This work is licensed under a
  \href{https://creativecommons.org/licenses/by/4.0/}{Creative Commons
  Attribution 4.0 International License (CC BY 4.0)}.}
\par\addvspace{\medskipamount}{\footnotesize\noindent\preprintlicense\par}

\section{Introduction}\label{sec:introduction}

Forests play a critical role in the global carbon cycle, biodiversity conservation, and
the provision of ecosystem services. Accurately modeling forest growth and dynamics is
therefore essential for assessing the effects of climate change, supporting sustainable
forest management, and informing policy decisions. Process-based models such as 3-PG
simulate forest growth by representing key physiological processes and their responses to
environmental conditions \citep{Landsberg1997}.

The 3-PG model has been widely applied to simulate forest growth and productivity. It was
originally distributed as a spreadsheet-based tool and was later released as the R
package r3PG \citep{Trotsiuk2020b}. The package provides an R interface to a Fortran core
and is currently the most widely used \citep{Forrester2021a} and actively maintained  
implementation of the model. It supports single- and multi-stand simulations, sensitivity
analysis, and Bayesian calibration \citep{Trotsiuk2020b}. However, its architecture does
not support automatic differentiation, limiting its computational flexibility and
compatibility with gradient-based methods. As a result, Bayesian calibration using
gradient-free Markov chain Monte Carlo (MCMC) methods typically requires several hours to 
achieve convergence~\citep{Forrester2021a}. Because these methods cannot use local 
sensitivities to guide their exploration of the parameter space, they may require 
substantially more model evaluations than gradient-informed approaches to achieve a 
comparable fit.

Differentiable modelling offers a promising way to overcome these limitations and connect
process-based models with modern statistical and machine-learning methods. By expressing
a model as a differentiable function, derivatives of its outputs with respect to
parameters, inputs, and intermediate states can be computed automatically. This
capability enables efficient gradient-based calibration and optimization. It also allows
mechanistic models to be coupled with data-driven components and trained end to end while
retaining the established first principles. Such domain-informed and hybrid approaches
have already been successfully applied in biology \citep{AlQuraishi2021}, hydrology
\citep{Wang2024}, and the geosciences~\citep{Shen2023}. Differentiability can also
facilitate integration with approaches such as neural ordinary differential equations, in
which neural networks can be combined with differential-equation-based representations of
dynamical systems~\citep{Chen2018}.

Differentiable programming frameworks such as JAX~\citep{Bradbury2018} provide the
technical foundation for implementing these capabilities through automatic
differentiation, just-in-time (JIT) compilation and automatic vectorization~\citep{Kidger2022}. 
Automatic differentiation provides model gradients for gradient-based
inference and optimization, while JIT compilation can accelerate repeated model
evaluations. Together, these features can reduce both the computational cost and the
number of evaluations required for calibration. Reimplementing 3-PG in JAX therefore not
only improves computational efficiency but also provides a foundation for combining its
established process structure with data-driven components in future hybrid forest models.

In this paper, we present TrunX, a JAX porting of the 3-PG model that combines
differentiability with efficient compiled execution. We evaluate the implementation
through five case studies:

\begin{enumerate}
    \def\labelenumi{\arabic{enumi}.}
  \item
    A validation of our implementation against r3PG.
  \item
    A comparison of execution time against the reference implementation.
  \item
    A global sensitivity analysis used to select which parameters are worth calibrating.
  \item
    A calibration study that compares three strategies, two Bayesian and one gradient-based,
    against long-term forest monitoring data.
  \item
    A spatial simulation demonstrating automatic vectorization across model runs.
\end{enumerate}

\section{The 3PG model}\label{sec:3pg-model}

The 3-PG model uses species-specific parameters describing plant physiology and
morphology to simulate forest-stand development through five interacting submodels: light
interception, productivity, water balance, biomass allocation, and mortality
\citep{Landsberg1997}. These submodels update stand structure and biomass pools at
monthly time steps. Although 3-PG was originally developed for monospecific, even-aged
evergreen forests, it has since been extended to represent deciduous, uneven-aged, and
mixed-species stands \citep{Forrester2021b}.

The light sub-model calculates light absorption using species-specific light extinction
coefficients and leaf area index (LAI), with the horizontal canopy structure quantified
using fractional ground cover \citep{Forrester2014}.

The productivity submodel converts absorbed radiation into gross primary productivity
(GPP) using the species-specific canopy quantum efficiency, $\alpha_C$. Potential
productivity is reduced by modifiers representing temperature, frost, vapor pressure
deficit (VPD), soil moisture, soil fertility, atmospheric CO\textsubscript{2}, and stand
age. Net primary productivity (NPP) is then calculated as a constant fraction of GPP
\citep{Waring1998,SandsLandsberg2002}

The water sub-model calculates transpiration and soil evaporation using the
Penman--Monteith equation \citep{Monteith1965,Penman1948}. These are added to canopy
interception to predict evapotranspiration. Soil water is calculated as the difference
between evapotranspiration and rainfall, while draining off any water in excess of the
maximum soil water holding capacity \citep{Forrester2021b}.

The biomass allocation submodel distributes NPP among roots, stems, and foliage.
Allocation depends on environmental conditions, including soil fertility, VPD, and soil
moisture, while partitioning between stems and foliage also varies with tree size and age
\citep{Landsberg1997}.

Finally, the mortality sub-model calculates density-dependent mortality based on the
$-3/2$ self-thinning law \citep{Yoda1963}. After each time step, simulated biomass is
converted into output variables such as mean tree diameter, height, basal area, wood
volume, and biomass distributions using allometric relationships.

\section{Case study}\label{sec:case-study}

We re-implement all major published submodules and functions of 3-PG in JAX.
This implementation is particularly relevant for large-scale, spatial
simulations, as JAX enables efficient parallel computation and GPU acceleration. In
addition, its automatic differentiation capabilities allow gradient computations,
facilitating gradient-based parameter estimation, optimization, and sensitivity analyses.
Unlike previous implementations of 3-PG in Fortran~\citep{Adamsnd}, R~\citep{Trotsiuk2020b}, or JavaScript, 
our JAX implementation provides a state-of-the-art framework that supports both scalable
computation and differentiable model simulations.

\subsection{Implementation of 3PG model }\label{sec:implementation}

To illustrate the application of our 3-PG model, we use growth data from the Solling site
\citep{Reyer2020} in Germany, pre-processed by \citet{Trotsiuk2020b}. For most of the
analysis, except the spatial simulation presented in Section~\ref{sec:spatial}, we use the 
same Solling dataset. We validate our implementation against the Fortran-based r3PG package
\citep{Trotsiuk2020b} by running both implementations on Solling site data and comparing
the resulting monthly trajectories of diameter, height, basal area, and stem biomass.
Across all runs examined, the JAX and R implementations produce numerically equivalent
results, with differences limited to rounding error (see Fig.~\ref{fig:validation}).

\begin{articlefigure}[htb]
  \begin{center}
    \includegraphics[width=\linewidth]{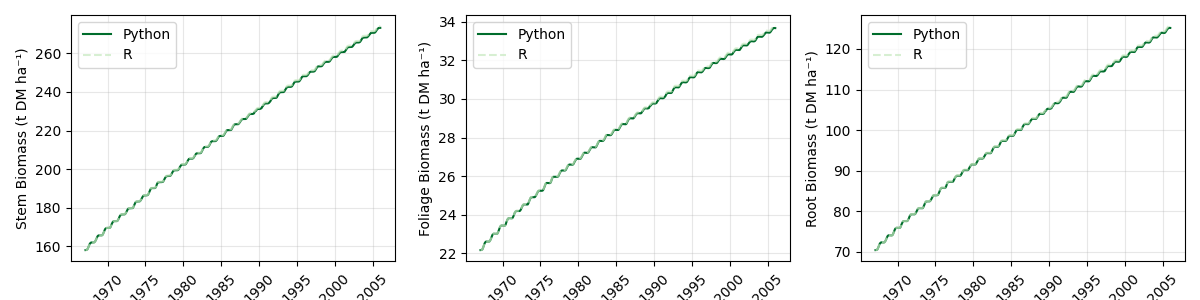}
  \end{center}
  \caption{3PG biomass prediction using our implementation (Python) and using r3PG
    reference implementation.\label{fig:validation}}
\end{articlefigure}

\subsection{JAX Performance and Scalability }\label{sec:performance}

To assess the computational performance of the JAX implementation, we compare its
execution time with the R implementation for different numbers of model runs
(Table~\ref{tab:performance}). The benchmarks are conducted on a Mac laptop equipped
with an Apple M3 processor and on a computational cluster equipped with Intel Xeon Gold
6542Y CPUs and an NVIDIA H100 NVL GPU. For a single run, JAX with CPU vectorization
(`vmap`) is already 3.6 times faster than the R implementation, whereas GPU execution
is slower due to the overhead associated with GPU initialization and data transfer.

As the number of runs increases, the advantages of JAX become substantially more pronounced.
For 1,000 runs, JAX achieves speedups of 27.5 times on the CPU and 138 times on the
GPU relative to R. For 100,000 runs, the JAX CPU implementation is 26.9 times faster,
while GPU execution achieves a speedup of approximately 4,878. These results demonstrate
the scalability of the JAX implementation and highlight the benefits of vectorized and
GPU-accelerated computation for large-scale 3-PG simulations.

\begin{articletable}[!ht]
  \centering
  \caption{Performance comparison between R and JAX implementations (on GPU and local
  laptop).\label{tab:performance}}
  \begin{tabular*}{\textwidth}{@{\extracolsep\fill}lccccc@{\extracolsep\fill}}
    \toprule
    \textbf{\# runs} & \textbf{R (s)} & \textbf{JAX CPU (s)} & \textbf{JAX GPU (s)} &
    \textbf{JAX CPU vs R} & \textbf{JAX GPU vs R} \\
    \midrule
    1 & 0.017 & 0.005 & 0.050 & 3.6$\times$ & 0.36$\times$ \\
    1,000 & 5.5 & 0.2 & 0.040 & 27.5$\times$ & 138$\times$ \\
    100,000 & 375.6 & 13.98 & 0.077 & 26.9$\times$ & 4,878$\times$ \\
    \bottomrule
  \end{tabular*}
\end{articletable}

\subsection{Sensitivity Analysis}\label{sec:sensitivity}

In our third case study, we conduct a global sensitivity analysis of the 3-PG model
parameters to identify which parameters are most influential and therefore potentially
worth calibrating against observations. We use the Morris method \citep{Morris1991}, as
implemented in SALib \citep{Iwanaga2022}. This method perturbs one parameter at a time
along a set of trajectories through parameter space and summarizes parameter influence
using the mean absolute elementary effect, $\mu^*$, on a given model output.

The computational efficiency of the JAX implementation makes it possible to perform this
analysis at a relatively low computational cost. By JIT-compiling the model and batching 
model evaluations where appropriate, we evaluated 1,000 Morris trajectories with 
20 discretization levels per parameter in approximately 6 seconds, whereas the 
equivalent analysis using r3PG required more than 11 minutes on a local laptop.This
illustrates the advantage of the JAX implementation for computationally intensive
analyses requiring large numbers of repeated model evaluations.

We screen all physiological model parameters together with the observation-error terms
associated with each output variable. Parameters are ranked according to their $\mu^*$
values for the model outputs used in the calibration, namely diameter, stem, root and
foliage biomass, height, and basal area. We use the same dataset that is used in
Section~\ref{sec:implementation}.

It is important to note that only diameter and height are directly observed in the field.
The remaining outputs, biomass pools, and basal areas are derived from these observations
through allometric relationships. Consequently, their apparent sensitivity reflects not
only the sensitivity of the underlying growth dynamics but also the sensitivity of the
allometric relationships used to derive these quantities.

The sensitivity analysis identifies the parameters with the greatest influence on each
output variable. Overall, the parameters identified as most sensitive based on their
total contribution to the log-likelihood across output variables are consistent with the
findings of \citep{Trotsiuk2020b}.

\begin{articlefigure}[htb]
  \begin{center}
    \includegraphics[width=\linewidth]{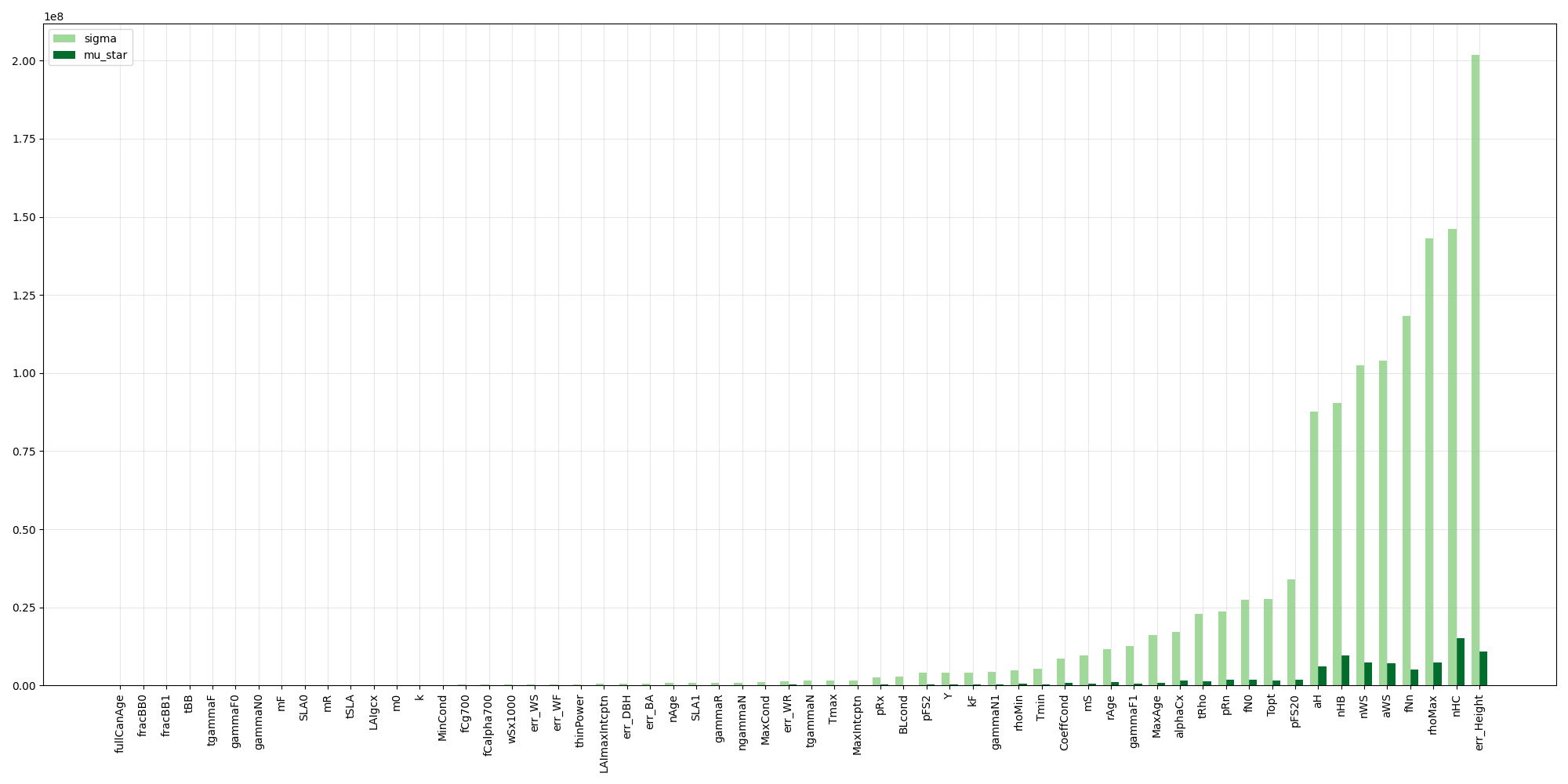}
  \end{center}
  \caption{Morris sensitivity analysis on Solling data Germany.\label{fig:sensitivity}}
\end{articlefigure}

\subsection{Bayesian calibrations and gradient descent }\label{sec:calibration}

In the fourth case study, we calibrate the model for the Solling site against biomass
observations from the dataset employed in the sensitivity analysis. The calibration
includes the 20 physiological parameters and observation-error term for the biomass data.

Hereafter, we compare one point-estimation method with two probability-based calibration methods. 
For point estimation, we use gradient-based optimization with the Adam algorithm and with a learning rate of $10^{-3}$.
JAX uses automatic differentiation to compute gradients of model outputs with respect to model parameters. 
An optimizer then uses these gradients to identify parameter values that minimize the discrepancy between predictions and observations. 
This approach produces point estimates but does not sample from a posterior distribution and therefore does not directly quantify parameter uncertainty. 
It is not supported by the reference r3PG implementation because r3PG does not expose model gradients.

For probability-based estimation, we perform Bayesian calibration using two samplers: DEMetropolis and the No-U-Turn Sampler (NUTS). DEMetropolis enables a direct comparison with the r3PG implementation \citep{Forrester2021b,Trotsiuk2020a}. 
NUTS is a gradient-based Bayesian sampler that can converge more rapidly. 
Unlike gradient-based optimization, both Bayesian approaches estimate posterior distributions and thus quantify parameter uncertainty.

Computational costs varies substantially among the three calibration approaches. The
gradient-free DEMetropolis sampler required approximately 17 hours to converge,
whereas the gradient-informed No-U-Turn Sampler (NUTS) converged in just over three
hours. Gradient-descent optimization is considerably faster, reaching convergence within
minutes. Model performance is assessed using the root mean squared error (RMSE) between
observed and predicted foliage, stem, and root biomass; the results are reported in
Table~\ref{tab:rmse}. NUTS achieves RMSE values comparable to those of DEMetropolis in
substantially less time. These results demonstrate a key advantage of the JAX
implementation: by making 3-PG differentiable, it supports both more efficient
gradient-informed Bayesian sampling and rapid gradient-based parameter optimization.

\begin{articletable}[!ht]
  \centering
  \caption{RMSE of output variables against the observed Solling site data, using the
    default parameter values, the PyMC (DEMetropolis) calibration, and the gradient-based
  calibration approach.\label{tab:rmse}}
  \begin{tabular*}{\textwidth}{@{\extracolsep\fill}lccccc@{\extracolsep\fill}}
    \toprule
    \textbf{Metrics} & \textbf{Variable} & \textbf{Default} & \textbf{DEMetropolis} &
    \textbf{NUTS} & \textbf{Gradient descent} \\
    \midrule
    RMSE & Foliage biomass (tDM per hectare) & 2.86 & 0.42 & 0.42 & 0.43 \\
    RMSE & Stem biomass (tDM per hectare) & 20.83 & 3.15 & 3.15 & 3.25 \\
    RMSE & Root biomass (tDM per hectare) & 9.25 & 1.43 & 1.43 & 1.5 \\
    \multicolumn{2}{l}{Number of model evaluations (forward passes)} & 1 & 6000000 & 2000 & 2500 \\
    \multicolumn{2}{l}{Wall-clock time (hours)} & $1.4 \times 10^{-4}$ & 17.33 & 3.03 &
    $1.4 \times 10^{-3}$ \\
    \bottomrule
  \end{tabular*}
\end{articletable}

To evaluate the uncertainty associated with the Bayesian calibration, we draw 500 samples
from the posterior parameter distribution and use these samples to generate posterior
predictive trajectories. The resulting 95\% credible intervals are shown in
Fig.~\ref{fig:calibration}, together with the model fits obtained using the
gradient-based calibration. This comparison illustrates the agreement between the
sampling-based and gradient-based calibration approaches while highlighting the
substantially lower computational cost of the latter.

\begin{articlefigure}[htb]
  \begin{center}
    \includegraphics[width=\linewidth]{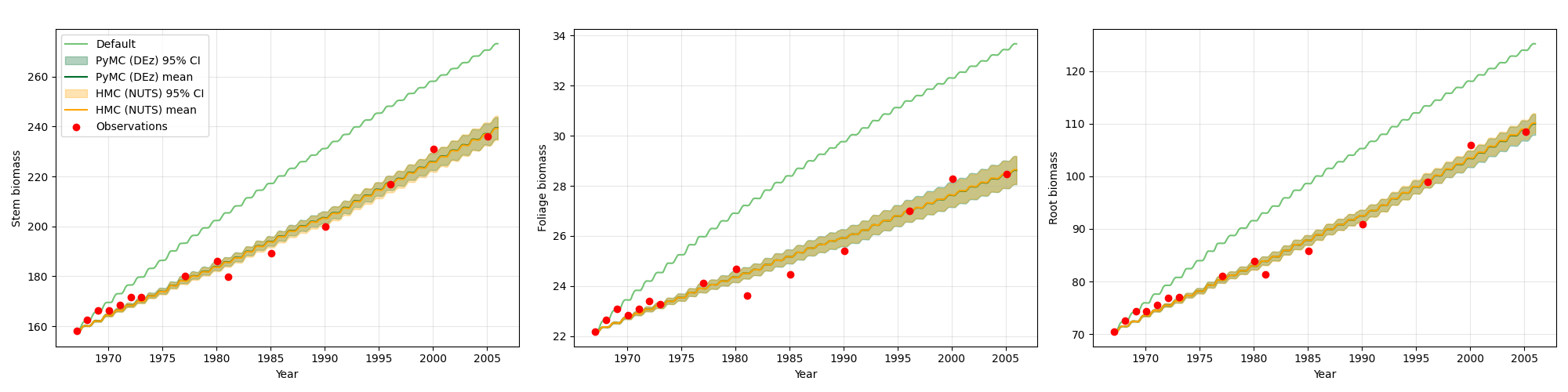}
  \end{center}
  \caption{Posterior predictive trajectories and 95\% credible intervals based on 500
    samples from the Bayesian posterior distribution using DEMetropolis and NUTS, along
    with predictions from the gradient-descent calibration and default
  parameters.\label{fig:calibration}}
\end{articlefigure}

\subsection{Spatial simulation}\label{sec:spatial}

For our final case study, we use the spatial input data prepared by~\citet{Trotsiuk2020b} 
to simulate Picea abies stand biomass across the forested area of Switzerland. The dataset
provides climate and soil information at a 1~$\times$~1~km spatial resolution, resulting
in 18,745 grid cells across Switzerland. The climate data
consist of interpolated meteorological variables that are reported to be derived from
MeteoSwiss observations, and soil type and available soil water from European Soil
Database Derived data. We use these spatially explicit inputs to initialize independent
3-PG simulations for each grid cell. Stands are initialized as 2-year-old plantations
with an initial density of 2,500 trees per hectare and simulated until age 30 under the
average climate conditions observed during 1961-1990.

To quantify uncertainty in the spatial predictions, we draw 500 parameter sets from the
posterior distribution obtained in the previous MCMC calibration and run the model for
each parameter combination across all grid cells. 
Runs were automatically vectorized along the parameters and locations dimensions using 
JAX built-in functions transformations (vmap).
The simulation, which was previously
reported by~\citet{Trotsiuk2020b} to require approximately 1 hour on a cluster with 50 processes,
completes in less than 15 seconds using our JAX implementation on a single NVIDIA
H100 NVL GPU, i.e. approximately 240$\times$ faster. This substantial reduction in
execution time demonstrates the ability of the JAX implementation to efficiently perform
large-scale spatial simulation.

\begin{articlefigure}[htb]
  \begin{center}
    \includegraphics[width=\linewidth]{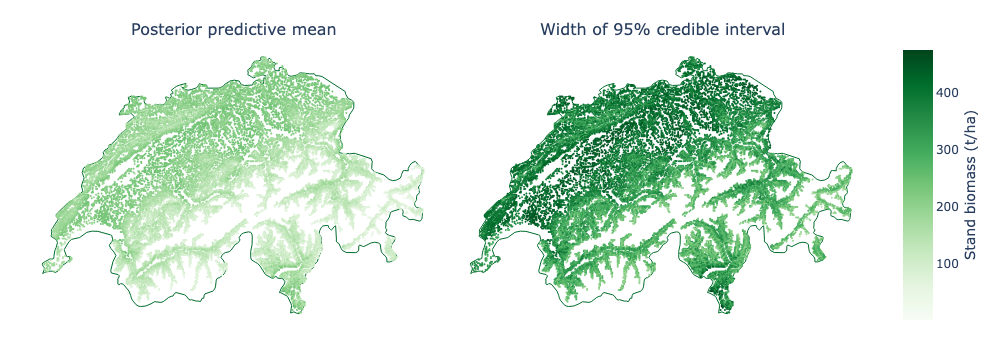}
  \end{center}
  \caption{Average and credible interval of stand biomass on 18,745 grid points across
  Switzerland.\label{fig:spatial}}
\end{articlefigure}

\section*{Conclusions}\label{sec:conclusions}

We developed a JAX implementation of 3-PG that substantially reduces execution time for
computationally intensive applications. JIT compilation and vectorization accelerated
repeated simulations on CPUs, while GPU execution delivered particularly large gains for
global sensitivity analysis, Bayesian calibrations, and national-scale spatial
simulation. These improvements make analyses involving millions of model evaluations
feasible within substantially shorter time frames.

Automatic differentiation is a second major contribution of the implementation. By
providing gradients of calibration objectives with respect to model parameters, it
enables gradient-informed Bayesian sampling and direct gradient-based optimization. In
our case study, gradient-based optimization achieved predictive errors comparable to the
MCMC point estimate within minutes rather than hours. Differentiability therefore expands
the methodological capabilities of 3-PG beyond conventional gradient-free calibration and
creates opportunities for efficient multi-site calibration, data assimilation, and hybrid
process-based and machine-learning models.

The JAX implementation produced numerical results consistent with the established r3PG
implementation for the evaluated simulations. Overall, the combination of faster
execution and automatic differentiation provides a scalable framework for forest-growth
calibration, uncertainty analysis, and spatial simulation from individual stands to large
landscapes.

\backmattersection{Funding}
Swiss National Science Foundation (SNSF), Grant/Award Number: 237270; ``Domain-Informed System Dynamics Modelling of Tree Growth and Mortality under Changing Climatic Conditions''

\backmattersection{Conflict of Interest Statement}
The authors declare no conflicts of interest.


\backmattersection{Acknowlegments}
The authors thank Dr. Fabian Bernhard (University of Bern), Dr. Arthur Gessler, Dr. Mirko Lukovic, and Dr. Volodymyr Trotsiuk (Swiss Federal Institute for Forest, Snow and Landscape Research) for their valuable support and helpful discussions.

\backmattersection{Data Availability Statement}
The code supporting this study is openly available on GitHub at https://github.com/simlab-vs/TrunX and archived on Zenodo at https://doi.org/10.5281/zenodo.22097261.

\newpage
\printbibliography

\end{document}